# Encrypted Federated Search Using Homomorphic Encryption

Om Rathod[1] , Aastha Baid[1], Aswani Kumar Cherukuri[1*]
[1]School of Computer Science Engineering and Information Systems,

Vellore Institute of Technology, Vellore India.

*cherukuri@acm.org

# Abstract

The sharing of information between agencies is effective in dealing with cross-jurisdictional criminal activities; however, such sharing is often restricted due to concerns about data privacy, ownership, and compliance. Towards this end, this work has introduced a privacy-preserving federated search system that allows law enforcement agencies to conduct queries on encrypted criminal databases by utilizing Homomorphic Encryption (HE). The key innovation here is the ability to execute encrypted queries across distributed databases, without the decryption of the data, thus preserving end-to-end confidentiality. In essence, this approach meets stringent privacy requirements in the interests of national security and regulatory compliance. The system incorporates the CKKS and BFV scheme embedded within TenSEAL, with each agency holding its key pair in a centralized key management table. In this federated search, encrypted queries are computed on the server side, and only authorized clients can decrypt the computed results. The matching of agencies is flexible for working in real-time while at the same time being secure and scalable while preserving control over data and the integrity of the process. Experimental results demonstrate the model. This paper also provide the implementation code and other details.

# 1. Introduction

In the current scenario, digital data proliferates exponentially, and avowedly rapid internet technology advancements have meant that more cybersecurity threats hit sensitive government and law enforcement operations than any other domain. Cybercrime and digital intrusion continue to rise internationally, which has made it increasingly pressing for the safe handling, sharing, and query of confidential data. Law enforcement agencies find themselves teetering on very major lines between collaboration and, at the same time, protecting the confidentiality and integrity of sensitive criminal information. Modern criminal investigations tend to span jurisdictions and involve several local, national, and international law enforcement agencies. Every agency maintains its database of records, usually for internal access and entries aligned to regional laws on data protection. These databases are well secured within their confines but end up being drastically obstructive when it comes to expedited access to essential information during cross-jurisdictional investigations. The inability to carry out federated, real-time searches among the databases delays investigations, weakens coordination between agencies, and finally compromises national security and public safety.

Federated search technology promises a rewriting of the common rules of querying across decentralized data sources without aggregating them. However, persistent challenges remained. Privacy concerns loom large, with agencies not comfortable sharing any of their raw or decrypted data for fear of unauthorized access, surveillance, or misuse. Legal and jurisdictional constraints vary from region to region, complicating compliance with agreements on data sharing. Most traditional federated approaches also divulged metadata or content of the query, which are more easily targeted for attacks, thus increasing the risk of leak or breach.

This work now proposes a secure federated querying framework based on HE to overcome such critical limitations. Direct computation on the encrypted form of the data is possible using HE. It ensures that there is still additional hiding of raw data during processing. Hence, HE is very appropriate for secure collaboration between agencies in the law enforcement and intelligence world. In precisely the same way, even agencies without decrypting private information on the server side can conduct keyword-based queries, similarity searches, or pattern-matching operations on records that are encrypted. In this research, we deploy secured filtering for maintaining end-to-end confidentiality of data, using a dot product operation to match encrypted query vectors against encrypted records with a focus on the "crime type" column.

## 2. Background on Homomorphic Encryption

Cryptography is the mainstay of our modern secure communication and data management systems. It deals with the conversion of information into an unreadable format (wherein such information is transformed through decryption protocols only by authorized users); symmetric-key cryptography uses a single counterpart shared key, while asymmetric-key cryptography employs a public-private key pair. The asymmetric machinery is the underlying principle of HE, a scheme that allows operations to be performed on encrypted data without requiring access to plaintext.

Based broadly on the number of operations, HE comprises partially homomorphic encryption (PHE), somewhat homomorphic encryption (SHE), and fully homomorphic encryption (FHE). FHE allows both addition and multiplication operations and can express arbitrary computations over ciphertexts. Libraries notable for implementing HE include Microsoft SEAL, which supports both the BFV (Brakerski/Fan-Vercauteren) scheme for exact integer arithmetic [1] the and CKKS (Cheon-Kim-Kim-Song) scheme for approximate arithmetic on real numbers [2]. TenSEAL, based on Microsoft SEAL, extends this functionality for encrypted tensors, thereby supporting vectorized operations like dot product and element-wise multiplication that are critical for search and machine learning [3].

BFV can apply operations to encrypted integers, making it suitable to encrypt sensitive fields containing names, locations, and timestamps. CKKS, on the other hand, presents an approximate operation on real numbers and is applied to encode numeric representations of categorical or continuous features. The combination of these two schemes permits the use of hybrid encryption architectures allowing optimization of performance and precision for different data types in one system. The dot product operation that underpins the comparison between encrypted search vectors and encrypted database rows in this work is implemented under CKKS in support of privacy-preserving matches in federated search.

### 2.1 Law Enforcement Context and Motivation

The enforcement of laws by a LEA occurs within a defined jurisdiction. Local crime enforcement is carried out by agencies in coordination with others, such as city police departments, federal intelligence units, and intergovernmental organizations, to prevent crimes ranging from local offenses to international terrorism. Their databases typically contain a lot of information concerning criminal incidents, the list of suspects, the timeline of events, forensic evidence, and surveillance. Because of their sensitive nature, these databases serve as prime targets for cyber adversaries and therefore must be protected through stringent access control and encryption.

Real-time collaboration across LEAs has in recent times become all the more pressing due to cross-border crimes, organized cybercrime rings, and the worldwide nature of threats. Unfortunately, the existing data-sharing infrastructure is inadequate to support such collaboration. Away from the question, direct election by agencies of data-sharing violates privacy statutes and leads such agencies into legal liabilities or possible political scrutiny. Furthermore, traditional cryptographic solutions lack the capability for secure querying and computation over encrypted records.

## 2.2 Objectives of the Work

This research is involved in designing and building a privacy-preserving federated querying system with the aim of inter-agency cooperation with the following cardinal objectives:

1. Perform secure queries over encrypted law enforcement databases using a combination of BFV and CKKS HE schemes.
2. Allow federated visualization in a way in which the data stays with the originating agency but is still usable in global queries.
3. Protecting privacy and confidentiality are the means that prevent the leaking of sensitive information whether be the process of search, computation, or transmission.
4. Supporting real-time operations where encrypted search is fast and scalable to accommodate large datasets across multiple nodes.
5. Construct an extensive cryptographic framework that will allow enriching the functionalities to include things like secure machine learning keyword search or privacy-preserving analytics in the foreseeable future.
6. Encrypted dot product-matching possibly between the query vector and the entries of the database-related crime class enables secure, efficient, and relevant record retrieval.

Through the injection of homomorphic encryption into the federated database querying process, this research contributes to the establishment of a secure and scalable infrastructure for inter-data collaboration in law enforcement that leads to the fast-tracking of crime detection, enhanced intelligence sharing, and generally better standards of data security.

# 3. Background and Motivation

## 3.1 Federated Search Systems

Federated search is the process of querying multiple decentralized databases without the need to consolidate all the data in a single repository. This is how large-scale systems like healthcare, finance, or academic researchers can retain their autonomy over their data while collaborating beyond institutional boundaries. Such systems mainly rely on middleware to aggregate and dispatch the queries to the endpoints, which return the results in a unified way [4].

In a law enforcement context, however, federated search systems face a lot more scrutiny. The data such as criminal records, surveillance logs, and intelligence records are extremely sensitive, thus requiring much tighter controls of confidentiality. Things such as query keywords, frequency, or origin agency become metadata that conventional federated systems generally expose and may be able to be used for inference attacks by adversaries. Moreover, the very assumptions of trust embedded within such models tend to be untenable in law enforcement given that agencies may be hamstrung by national security edicts, laws, or political boundaries from cooperating through these typical federated systems. Therefore, by their nature, they are unable to provide much in the way of privacy-preserving mechanisms which are just what is needed for many multi-agency criminal investigations [5].

## 3.2 Homomorphic Encryption for Secure Processing

HE changes the context of computing securely by making it possible to execute mathematical operations directly on encrypted data without requiring decryption at any stage. Therefore, this property makes HE an attractive solution to problems where privacy consideration must be satisfied alongside active operational utility [6].

The level of operations that can be performed therefore puts HE schemes in classification. Partially Homomorphic Encryption (PHE) supports addition or multiplication. Somewhat Homomorphic Encryption (SHE) permits only a limited number of both operations while expressing, for any given number of operations over ciphertext, Fully Homomorphic Encryption (FHE). We take the BFV and CKKS schemes from Microsoft SEAL implementation for this project. BFV is more appropriate for doing exact computations on encrypted integers, supporting the matching of structured fields such as names or IDs. Whereas, CKKS allows approximate arithmetic over real-valued vectors and so it is perfect for high dimensional, floating-point encoded representations of synonyms like crime types [7].

CKKS enables all queries as well as records stored in associated encrypted databases to be encrypted. This allows the user to perform dot-products to compute encrypted similarity scores for the system to recover the most relevant documents based on matching encrypted crime categories without ever revealing raw data in the server. Jain and Cherukuri [17] have provided a detailed analysis of FHE schemes. Recently Adamsetty et al., [18] have analysed the performance of machine learning classification models accuracy on HME data. Savadatti et al., [19] have analysed the quantum FHE schemes.

## 3.3 Why Encrypted Federated Search?

In sensitive data-sharing environments, the combination of federated search and HE addresses three foundational concerns: trust minimization, auditability, and regulatory compliance. Encrypted federated search allows agencies to collaborate without trusting each other with raw data or central authority. Since all calculations are performed over ciphertext, even the party submitting the query cannot ascertain the plaintext data from another agency. This model guarantees confidentiality under almost adversarial or semi-trusted scenarios, which is an obligatory feature for inter-governmental collaborations [8].

Moreover, this decentralized system provides a means for auditability. Query logging, access patterns, and data usage can all be cryptographically verified without exposure to their content. These features thus propagate transparent oversight while remaining compliant with regulations like GDPR or CJIS, which tightly regulate data sovereignty and access control [9]. Finally, the implemented system guarantees that queries remain riverine to local legal restriction, whereby data ever left in agency custody is not readable. The presented design thus juxtaposes operational agility against high-principled assurance in privacy; hence making it an attractive framework for modern law enforcement collaborations.

# 4. Related Work

Federated search, a methodology enabling unified querying across distributed, autonomous datasets, has seen increasing adoption in sectors where centralizing sensitive data is infeasible. Unlike federated learning, which primarily addresses collaborative model training, federated search facilitates data discovery without compromising raw data privacy. In this section, we review previous work on secure federated querying, HE in cross-agency systems, and the limits of traditional privacy-preserving mechanisms.

## 4.1 Secure Federated Search Systems

A growing body of research focuses on secure federated querying where encrypted datasets are queried without decryption. Chen et al. [10] proposed a privacy-preserving federated search system across distributed datasets using searchable encryption and attribute-based access control. Their system supports secure keyword queries but lacks full homomorphic processing, restricting the scope to exact keyword matching. Similarly, Li et al. [11] introduced a decentralized privacy-preserving system for healthcare federated search using proxy re-encryption, which allowed limited search flexibility and was not resilient to access-pattern leakage.

While these works contributed to federated query execution, they often rely on pre-processing indices or secure hardware and do not achieve full homomorphic encrypted computation over query vectors and record sets. Most systems require partial trust in aggregators or intermediaries and do not preserve full confidentiality during computation.

## 4.2 Homomorphic Encryption in Federated Contexts

HE has become central to privacy-preserving computation. Fully homomorphic encryption (FHE), introduced by Gentry in 2009, enables arbitrary computation on encrypted data. In practice, leveled or somewhat homomorphic encryption is preferred for efficiency. Microsoft SEAL's CKKS scheme [12] supports approximate arithmetic, making it suitable for encrypted vector operations like dot pro and ducts, ideal for search matching, while BFV [13] offers exact arithmetic for categorical fields such as names or crime types.

Recent work by Li et al. [14] explores secure information retrieval using homomorphic dot products over encrypted data. They demonstrate that CKKS enables accurate similarity search on encrypted feature vectors with acceptable computational cost. Another approach by Wang et al. [15] employs hybrid HE for encrypted facial recognition, combining BFV and CKKS schemes for different data types, achieving both precision and confidentiality.

These frameworks, however, have not been implemented in federated law enforcement environments where cross-jurisdictional search and compliance requirements must be considered.

## 4.3 Limitations of Traditional Privacy Methods

Conventional cryptographic techniques such as Secure Multi-Party Computation (SMPC) and Differential Privacy (DP) offer theoretical privacy guarantees but are rarely scalable in operational contexts. SMPC protocols often incur high communication and computation overhead, unsuitable for real-time multi-database querying [16]. Differential privacy, while useful in statistical analysis, introduces noise that can degrade query accuracy, especially in forensic and intelligence workflows that require high fidelity. Furthermore, these methods do not allow meaningful encrypted computation beyond fixed query formats or statistical summaries and often require all parties to be online synchronously or to pre-define the computation format.

## 4.4 Gap Identification

While previous systems support aspects of encrypted search or federated query processing, few address the full stack required for privacy-preserving collaboration in law enforcement. Existing literature lacks a comprehensive framework where:

1. Multiple autonomous law enforcement agencies retain control over their encrypted databases,
2. Secure querying is achieved using homomorphic encryption (not trusted intermediaries),
3. Different data types (numerical and categorical) are supported via CKKS and BFV hybrid encryption,
4. Real-time vector-based searches are possible with no plaintext exposure at any stage.

Our work fills this gap by proposing a federated encrypted search architecture combining CKKS (for encrypted similarity search over real-valued encodings of features) and BFV (for exact matching of sensitive identifiers). We also address key separation, decentralized key storage, and secure matching using encrypted dot products—all within a framework compatible with legal and regulatory compliance constraints. Finally, table 1 Brings a comparison between the other related works and the work done by us in this paper.

| Paper / Work | Domain | Encryption Type | Federated Search | Search Over Encrypted Data | Key Contributions |
|---|---|---|---|---|---|
| Chen et al. (2021) [6] | Cross-agency Search | Searchable Encryption | Yes | Keyword-based only | Privacy-preserving federated query using attribute-based access control |
| Li et al. (2020) [7] | Healthcare | Proxy Re-Encryption | Yes | Limited | Cross-institution privacy-preserving search for EHRs |
| Li et al. (2022) [8] | Vector Search | CKKS | No | Approximate vector search | Dot product-based search over encrypted vectors |
| Wang et al. (2023) [9] | Biometric Search | BFV + CKKS (Hybrid HE) | No | Face recognition search | Hybrid encryption for privacy-preserving search on sensitive and numerical data |
| Our Work | Law Enforcement Federated Search | BFV + CKKS (Hybrid HE) | Yes | Dot-product similarity over encrypted data | End-to-end encrypted federated querying for criminal records across multi-agency databases |

**Table 1. Comparative Summary of Prior Work**

# 5. System Architecture and Design

## 5.1 Overview of Architecture

The proposed system architecture implements a secure multi-agency criminal database search framework utilizing homomorphic encryption techniques. Figure 1. represents how multiple participating law enforcement agencies, each maintaining their own encrypted criminal records database, are interconnected through a secure query processing infrastructure. This design enables cross-agency search capabilities while preserving the confidentiality of both the search queries and the stored criminal records. The architecture employs a hybrid encryption approach utilizing both BFV and CKKS homomorphic encryption schemes. This hybrid approach allows for exact integer-based encryption of textual data via BFV while enabling approximate floating-point operations necessary for vector similarity searches via CKKS. A centralized key management system maintains separation between encryption keys and encrypted data, strengthening the security posture of the overall system.

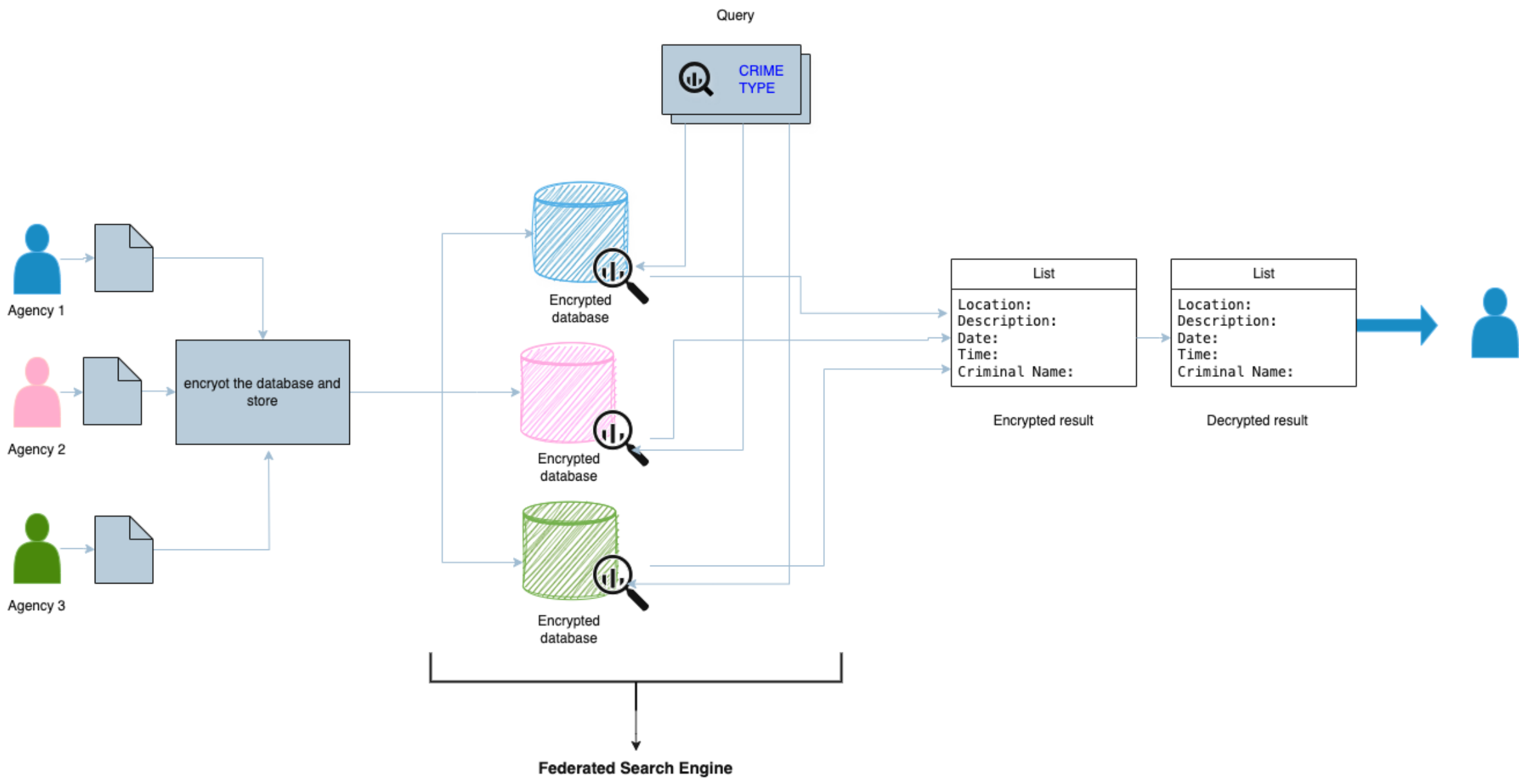

*Figure 1. System Architecture*

## 5.2 Key Components

The architecture of the system is the definition of various treatments that are interconnected yet perform different functions in the secure search framework as follows:

Query Encryption Module: It is responsible for changing the plaintext terms of the search query into homomorphic encrypted representations. The search word terms are vectorized and encrypted using the CKKS scheme to facilitate similarity-based comparison of the encrypted data. The module connects with the central key management system to gain access to the required public encryption keys.

Secure Query Dispatcher: It is responsible for the secure incorporation of encrypted queries the with member agency database. It contains connectivity parameters for the entire registered agency database and manages the necessary authentication credentials for cross-agency communication. The dispatcher applies networking security protocols to protect in-transit query data.

Homomorphic Search Engine: This function is performed after the encrypted search on encrypted criminal records databases. Two types of searches are carried out in this: vectors are compared for exact matches, and dot product computation for similarity-based searches. No decryption of the sensitive criminal data fields is needed to process encrypted CKKS vectors by the engine.

Encrypted Result Aggregator: This collects and consolidates encrypted search results from varied agency databases. As a result, it maintains the privacy of results until authorized decryption is performed. The aggregator applies filtering and sorting operations on the encrypted result set to enhance the presentation.

Decryption Authority: This optional component gives approved access to open decrypted search results. It connects to the key management system to get a secret key for decryption. Access control limits this decryption of criminal records to selected authorized personnel.

Key Management System: A basic component in which the system is designed to host and distribute keys across the whole system is this one. It keeps BFV and CKKS keys locked down on centralized databases, disconnected from the encrypted criminal data. Access controls have been put on private keys, making public keys available for encryption operations.

## 5.3 Workflow Process

The sequential workflow of the system as depicted in Figure 2. ensures that data confidentiality is preserved during the search process:

The process kicks off when an investigator authorized by the agency submits a search through their interface. The search query typically contains details like crime type and may contain other search parameters. Query Encryption Module transforms this plaintext query into a vectorized representation compatible with encrypted comparison. In searches using crime type as input, the system one-hot encodes vectors for predefined crime categories.

The vectorized query is then encrypted using the CKKS homomorphic encryption scheme in which the context is made public from the Key Management System. The encrypted query thus keeps the semantic meaning of the search parameters but no unauthorized observer will ever know the content. The system, meanwhile, prepares for result decryption by making BFV contexts available.

The Secure Query Dispatcher distributes the encrypted query to all the agency databases switched on to configure the system. Each agency database stores its encrypted criminal records, encrypted through a hybrid model-BFV for textual fields (location, description, date, time, criminal name) and CKKS for searchable vector fields (crime type). This separation ensures optimal performance for different data types and continues to ensure security.

Receiving the encrypted query, each agency's Homomorphic Search Engine executes the search against its local encrypted database. For exact matching, the CKKS vectors stored on the client side are decrypted and compared to the search vector. For similar matching, the product is computed between the encrypted query vector from the client and the stored encrypted vectors. This similar result is identified with a concerning threshold value.

Upon detection of matches, the BFV encrypted records are retrieved and preserved in encrypted form. The encrypted results are collected by the Encrypted Result Aggregator while combining the results from all involved agencies while guaranteeing confidentiality. The aggregator may also filter and prioritize the encrypted data.

Finally, the matching records are decrypted by the Decryption Authority, applying the correct BFV private keys upon verification of the proper credentials. In this way, the records of criminal prosecution will be decrypted and human-readable, and the information of location, description, date, time, and criminal name is revealed only to personnel authorized to have it. The events of decryption are logged for auditing.

This workflow guarantees that criminal record data remains encrypted throughout the search process until the final authorized decryption step, thereby achieving the "searchable encryption with homomorphic decryption" paradigm, which is a balancing act between security requirements and useful search capabilities.

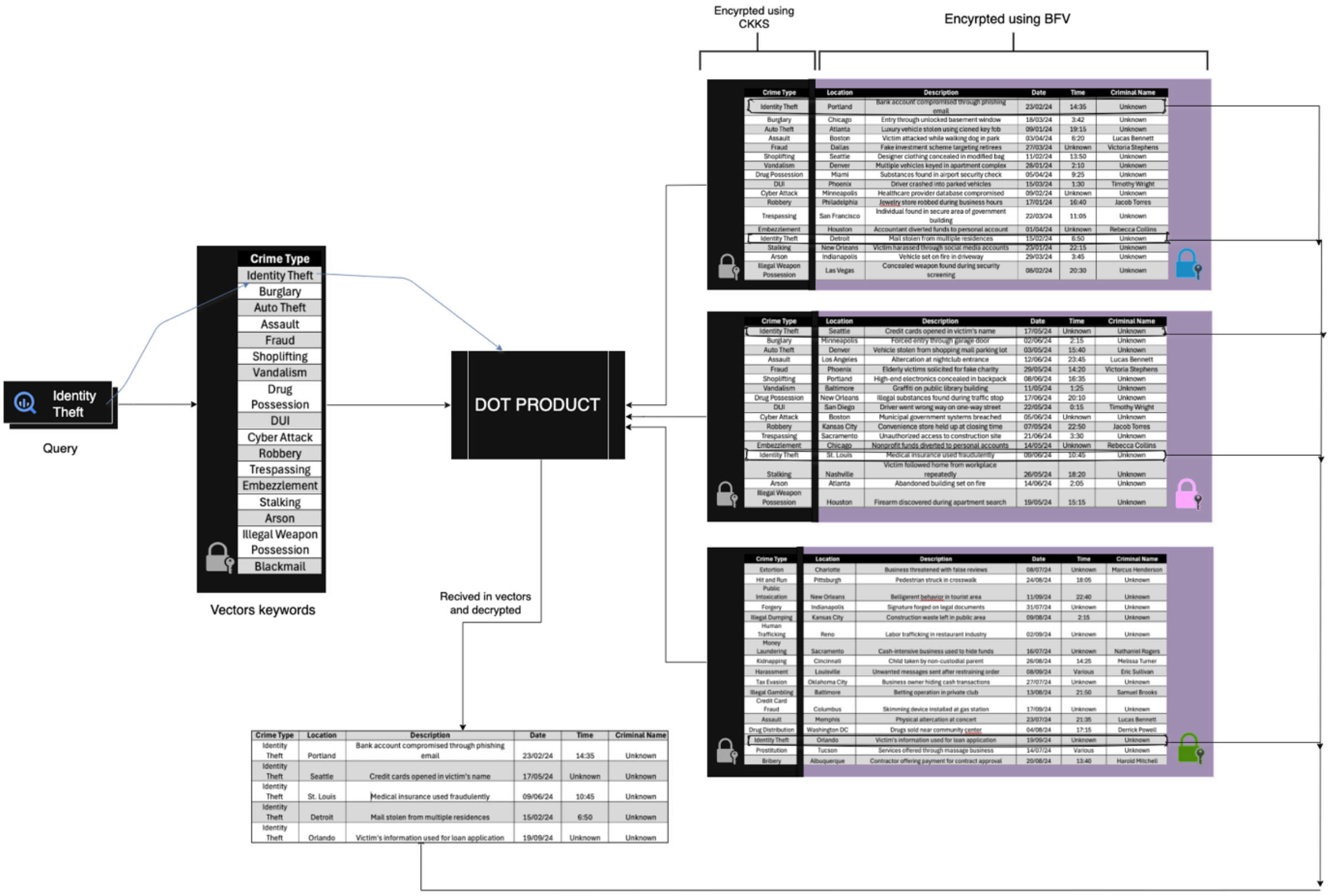

*Figure 2. Working of how the query is searched in the server.*

# 6. Implementation

## 6.1 Tools and Libraries

To implement our encryption and secure data storage system, we utilized a set of essential tools and libraries that support efficient development and integration with encryption techniques.

Programming Language

- Python
  Python's selection as the core programming language for this project is based on its simplicity and readability and because it has a strong presence in scientific and cryptographic libraries. Rapid Dev with integration capabilities with Machine Learning, Encryption, and database tools make it easy to work with.

Code Editor

- Visual Studio Code (VS Code)
  When we talk about code development, we refer to Visual Studio Code, an almost lightweight but powerful source code editor developed by Microsoft. It provided very good support for Python through extensions, debugging, integrated terminal, and Git. Its flexibility and user-friendly approach have made it an excellent editor for managing our codebase within the project.

Python Libraries

- TenSEAL

TenSEAL is a popular library in the world of homomorphic encryption operations on tensors designed specifically for machine learning and secure computing over encrypted data. In our project, we adopted TenSEAL for applying CKKS and BFV encryption schemes applying them in different domains of the crime report dataset. As a result, encrypted computations and searches can be conducted without exposing sensitive data.

```
1. pip install tenseal
```

- Psycopg2

  psycopg2 is a PostgreSQL adapter for Python that allows Python applications to interact with PostgreSQL databases securely and efficiently. The use of psycopg2 allowed us to manage database connections, execute SQL queries, and store or retrieve encrypted data from the database.

```
1. pip install psycopg2
```

## 6.2 Query Encoding and Encryption

This passage describes activities involved in constructing a search query used for privacy preserving extraction of encrypted crime reports. Tokenizing the search term, encoding it into a binary vector, and performing homomorphic evaluation using the CKKS encryption scheme provided by TenSEAL comprise the different stages of that procedure.

- Tokenization of search terms

  The search searches for a certain kind of crime tokenization that is processed, and the input term by which it is structured in a way as to enable encrypted vector comparison. Then the user-given search phrase would be matched with all the types of crimes that have been predefined. By this, the search term will always be "usually" consistent because it will match one among the categories present in the dataset used for vectors.

  In our implementation, tokenization would be achieved through a search term in all lowercase compared to all entries in the known kinds of crimes list. This would then yield a list of binary outcomes (0 or 1), which tells if the search term matched with a certain crime type.

  The list which was used for known crimes is given below:

```
1. Keywords = [ Identity Theft, Burglary, Auto Theft, Assault, Fraud, Shoplifting, Vandalism, Drug Possession, DUI, Cyber Attack, Robbery, Trespassing, Embezzlement, Stalking, Arson, Illegal Weapon Possession, Blackmail, Counterfeiting, Child Endangerment, Extortion, Hit and Run, Public Intoxication, Forgery, Illegal Dumping, Human Trafficking, Money Laundering, Kidnapping, Harassment, Tax Evasion, Illegal Gambling, Credit Card Fraud, Drug Distribution, Prostitution, Bribery, Insider Trading, Child Abuse, Grand Theft, Domestic Violence, Impersonation, Cyberbullying, Additional Insurance Fraud, Reckless Driving, Animal Cruelty, Carjacking, Online Scam, Ponzi Scheme, Theft of Services, Unlawful Possession of Firearm, Loitering, Check Fraud, Child Neglect, Corporate Espionage, Perjury, Drug Trafficking, Criminal Mischief, Disorderly Conduct, Larceny, Illegal Immigration, Homicide, Breaking and Entering, Copyright Infringement, Fare Evasion, Credit Card Theft, Illegal Firearms Sale, Pickpocketing, Child Pornography, Bomb Threat, Elder Abuse, Illegal Waste Disposal, Road Rage, Wire Fraud, Illegal Drug Manufacturing ]
```

- Binary vector encoding for encrypted search

The tokenized search term is converted into a binary vector, each element corresponding to a predefined crime category. The value 1 signals that the particular crime type applies to the search term, while a value of 0 signals that it does not. This binary representation underlies our encrypted search functionality.

The encoding takes place in the binary vector, which, under the CKKS homomorphic encryption scheme, is encrypted so that computations can occur directly on the encrypted data without decrypting it. In the project, the encoded search vector will be encrypted into a CKKS vector, where a scaling factor of 2^40 is to be used to keep precision in mathematical operations.

The binary encoding function is as follows:

```
def encode_search_term(search_term, known_crime_types):
        return [1 if search_term.lower() == crime.lower() else 0 for crime in known_crime_types]

```

The code which was used for encrypting with the CKKS encryption key:

```
encrypted_search_vector = ts.ckks_vector(ckks_context, encoded_search_vector, scale=2 ** 40)
```

- Homomorphic evaluation method (DOT PRODUCT)

In establishing a link between the encrypted query and the underground database entries, our system employs a homomorphic dot product operation - a cryptographic mathematical method allowing the similarity measurement of two vectors without disclosing them. More precisely, it operates over the encrypted search vector-the one representing the user query- and the crime type vector encrypted and called from the database. The start of this operation is followed by the decryption and decoding of this crime type stored vector before recoding and re-encrypting it to the CKKS format being used by the vector search. After this, both vectors multiply each element and then sum up their resulting values to find the dot product. The final decryption is carried out and applied to a threshold for establishing whether a meaningful match is present. This whole process takes place while holding data raw or sensitive within the highest secrecy.

```
decrypted_ckks_vector = decrypt_ckks_vector(ckks_context, encrypted_crime_type)
decoded_crime_type = [1 if val > 0.5 else 0 for val in decrypted_ckks_vector]
encrypted_search_vector = ts.ckks_vector(ckks_context, encoded_search_vector, scale=2 ** 40)
encrypted_db_vector = ts.ckks_vector(ckks_context, decoded_crime_type, scale=2 ** 40)
encrypted_product = encrypted_search_vector * encrypted_db_vector
dot_product_result = encrypted_product.sum()
dot_product_value = dot_product_result.decrypt()[0]
print("Dot Product:", dot_product_value)
match = dot_product_value > 0.01
```

## 6.3 Data Model

- Standardized criminal data format (simulated records)

To maintain consistency across all databases, a uniform schema has been implemented in each one. The structure ensures seamless integration of encrypted data and supports secure querying across multiple data sources. The schema for the encrypted_crime_reports table is defined as follows:

```
CREATE TABLE encrypted_crime_reports (
     id SERIAL PRIMARY KEY,  -- Auto-increment ID
     encryption_key_id INTEGER,
     crime_type BYTEA,  -- Encrypted crime type (as BYTEA)
     location BYTEA,  -- Encrypted location (as BYTEA)
     description BYTEA,  -- Encrypted description (as BYTEA)
     date BYTEA,  -- Encrypted date (as BYTEA)
     time BYTEA,  -- Encrypted time (as BYTEA)
     criminal_name BYTEA,  -- Encrypted criminal name (as BYTEA)
     encrypted_crime_type BYTEA  -- Encrypted vectorized crime type (as BYTEA)
 );
```

Simulation of records meant for the specific project in question. These records are standardized structures that can be easily extended or modified according to real-world datasets. All of the fields are stored in encrypted form in the form of homomorphic encryption schemes (CKKS or BFV) created as binary large objects (BYTEA). The encrypted_crime_type field contains the vectorized and encrypted version of the crime type, thus enabling fast and secure homomorphic search operations.

This guarantees standardization for encryption, indexing, and secure query processes across all connected databases. Below is a picture of what the database would look like and how encrypted values would be stored, providing a common foundation for secure multi-database search.

These records are specifically simulated for the project and represent a standardized structure that would allow adding to or modifying the schema for real-world datasets with ease. Each field is stored in encrypted form using homomorphic encryption schemes (CKKS or BFV), as binary large objects (BYTEA). The encrypted_crime_type field contains a vectorized and encrypted version of the crime type for fast and secure homomorphic search operations.

One of the guarantees of standardization is that encryption, indexing, and secure query processes are uniform across all connected databases. Figure 3. is a representation of what the database would look like and how the encrypted values would be stored so that it forms a common foundation for a secure multi-database search.

| | id [PK] integer | encryption_key_id integer | crime_type bytea | location bytea | description bytea | date bytea | time bytea | criminal_name bytea | encrypted_crime_type bytea |
|---|---|---|---|---|---|---|---|---|---|
| 1 | 1 | 2 | [binary data] | [binary d... | [binary data] | [binary d... | [binary d... | [binary data] | [binary data] |
| 2 | 2 | 2 | [binary data] | [binary d... | [binary data] | [binary d... | [binary d... | [binary data] | [binary data] |
| 3 | 3 | 2 | [binary data] | [binary d... | [binary data] | [binary d... | [binary d... | [binary data] | [binary data] |
| 4 | 4 | 2 | [binary data] | [binary d... | [binary data] | [binary d... | [binary d... | [binary data] | [binary data] |
| 5 | 5 | 2 | [binary data] | [binary d... | [binary data] | [binary d... | [binary d... | [binary data] | [binary data] |
| 6 | 6 | 2 | [binary data] | [binary d... | [binary data] | [binary d... | [binary d... | [binary data] | [binary data] |
| 7 | 7 | 2 | [binary data] | [binary d... | [binary data] | [binary d... | [binary d... | [binary data] | [binary data] |
| 8 | 8 | 2 | [binary data] | [binary d... | [binary data] | [binary d... | [binary d... | [binary data] | [binary data] |
| 9 | 9 | 2 | [binary data] | [binary d... | [binary data] | [binary d... | [binary d... | [binary data] | [binary data] |
| 10 | 10 | 2 | [binary data] | [binary d... | [binary data] | [binary d... | [binary d... | [binary data] | [binary data] |
| 11 | 11 | 2 | [binary data] | [binary d... | [binary data] | [binary d... | [binary d... | [binary data] | [binary data] |
| 12 | 12 | 2 | [binary data] | [binary d... | [binary data] | [binary d... | [binary d... | [binary data] | [binary data] |
| 13 | 13 | 2 | [binary data] | [binary d... | [binary data] | [binary d... | [binary d... | [binary data] | [binary data] |
| 14 | 14 | 2 | [binary data] | [binary d... | [binary data] | [binary d... | [binary d... | [binary data] | [binary data] |

*Figure 3. The data inside the database after encryption.*

- How datasets are encrypted and indexed

The hybrid encryption system protects the dataset by capitalizing on the strengths of CKKS and BFV homomorphic encryption schemes. With this scheme, we can have secure storage, as well as efficient encrypted search. Since this experiment was carried out under small conditions, we are simply using the vector version of the crime category for vector comparison.

1. Field-wise Hybrid Encryption

   Crime_Type:
   To facilitate an encrypted search, the crime_type field is first vectorized using a predetermined list of keywords as noted before. Each crime type is transformed into a binary vector indicating whether specific keywords are present or not.

   The vector is then encrypted using them based on the CKKS, which avails approximate arithmetic operations (like dot products) directly on encrypted data. The encrypted CKKS vector is stored twice:

   - Once as the main value of the crime_type field,
   - Again, in the encrypted_crime_type field for indexed, homomorphic search operations.

   Other Fields (like location, description, date, time, criminal name): These textual fields are encrypted individually using the BFV scheme, which is more suitable for an exact representation of data. It packs each character in the string into its ASCII value in a fixed-length vector. This vector is then encrypted and stored in the database as a BYTEA binary object.

2. Indexing for Secure Search

   To facilitate an efficient query operation over the encrypted data, a form of semantic indexing is constructed using the encrypted_crime_type field:

   - The encrypted crime-type vector acts as an encrypted index for the vectorized search through homomorphic operations.
   - During querying, the search term is vectorized in the same manner and encrypted under the same CKKS context.
   - The homomorphic dot product is then performed between the query vector and stored encrypted_crime_type vectors to determine how well they match without decrypting any data. Thus, privacy-preserving search.

   This indexing scheme does not require plaintext comparisons and ensures that sensitive crime-related data stays protected even when search operations are performed.

   The snippet of the entire code for the data encryption and data insertion is shown below.

```
import tenseal as ts
import psycopg2
import base64

# 🔐 Connect to PostgreSQL
conn = psycopg2.connect(
    dbname='NewDatabase',
    user='postgres',
    password='password',
    host='localhost',
    port='5432'
)

newdb = psycopg2.connect(
    dbname='Database 6',
```

```
    user='postgres',
    password='password',
    host='localhost',
    port='5432'
)

cur = conn.cursor()
newcur = newdb.cursor()

# 🔐 Step 1: Load BFV context from the database (for encrypting other fields)
cur.execute("SELECT public_key FROM encryption_keys WHERE id = %s", (4,)) # Adjust ID if needed
row = cur.fetchone()
if not row:
    raise Exception("No public key found with id = 3")

public_key_bytes = bytes(row[0])
# Load BFV context using the public key
bfv_context = ts.context_from(public_key_bytes)

# ✨ Vectorize crime type
def vectorize_crime_type(crime_type, keyword_list):
    vector = [0.0] * len(keyword_list)
    crime_type = crime_type.lower()
    for i, word in enumerate(keyword_list):
        if word.lower() in crime_type:
            vector[i] = 1.0
    return vector

# 🔒 Fixed-length encryption (BFV) for other fields
def encrypt_string_bfv(context, text, max_length=30):
    encoded = [ord(c) for c in text[:max_length]]
    encoded += [0] * (max_length - len(encoded))
    bfv_context = ts.bfv_vector(context, encoded)
    return bfv_context.serialize()

# 🔄 Encrypt and insert each report
for crime_report in crime_reports:
    # CKKS: Encrypt vectorized crime type
    vectorized = vectorize_crime_type(crime_report['crime_type'], keywords)

    # Ensure the vector is of type float (CKKS expects floats)
    vectorized = [float(v) for v in vectorized]

    # Encrypt with CKKS context
    ckks_vector = ts.ckks_vector(ckks_context, vectorized, scale=2 ** 40)
    encrypted_ckks = ckks_vector.serialize()

    # BFV: Encrypt other fields (location, description, etc.)
    encrypted_data = {
        key: encrypt_string_bfv(bfv_context, value)
        for key, value in crime_report.items() if key != 'crime_type'
    }

    # INSERT into encrypted DB
    newcur.execute("""
        INSERT INTO encrypted_crime_reports (
            encryption_key_id, crime_type, location, description,
            date, time, criminal_name, encrypted_crime_type
        ) VALUES (%s, %s, %s, %s, %s, %s, %s, %s)
    """, (
        4,
        psycopg2.Binary(encrypted_ckks), # vectorized crime_type (encrypted with CKKS)
        psycopg2.Binary(encrypted_data["location"]),
        psycopg2.Binary(encrypted_data["description"]),
        psycopg2.Binary(encrypted_data["date"]),
        psycopg2.Binary(encrypted_data["time"]),
        psycopg2.Binary(encrypted_data["criminal_name"]),
         psycopg2.Binary(encrypted_ckks) # storing the encrypted CKKS vector again for searching
    ))

# ✅ Finalize
```

```
conn.commit()
newdb.commit()
cur.close()
newcur.close()
conn.close()
newdb.close()

print("✅ Encrypted crime reports inserted using hybrid encryption (CKKS for crime type, BFV for others).")

```

## 641 Evaluation Parameters

- Query latency (vs. plaintext search)

To assess performance overheads imposed by our privacy-preserving search mechanism, we compared the query latency for traditional plaintext searches and our encrypted dot product method. In the plaintext case, the average query time for one PostgreSQL database (around 100 records) was 0.058 seconds represented in Figure 5; for three federated distributed databases, it was about 0.174 seconds. Using the homomorphic search, the cryptographic vector matching by way of dot product produced a return of approximately 20.5 seconds from the same three databases as shown in the Figure 4.

This entails considerable performance trade-offs:

Estimated Latency due to Encryption Search ≈ 20.5 - (3 × 0.058) ≈ 19 seconds

While the performance overhead is quite substantial, this is a necessary sacrifice to support secure, privacy-preserving querying over sensitive crime report data. It provides evidence of the obvious trade-off between speed and confidentiality that is concerned in the building of ever more privacy-centric systems based on homomorphic encryption

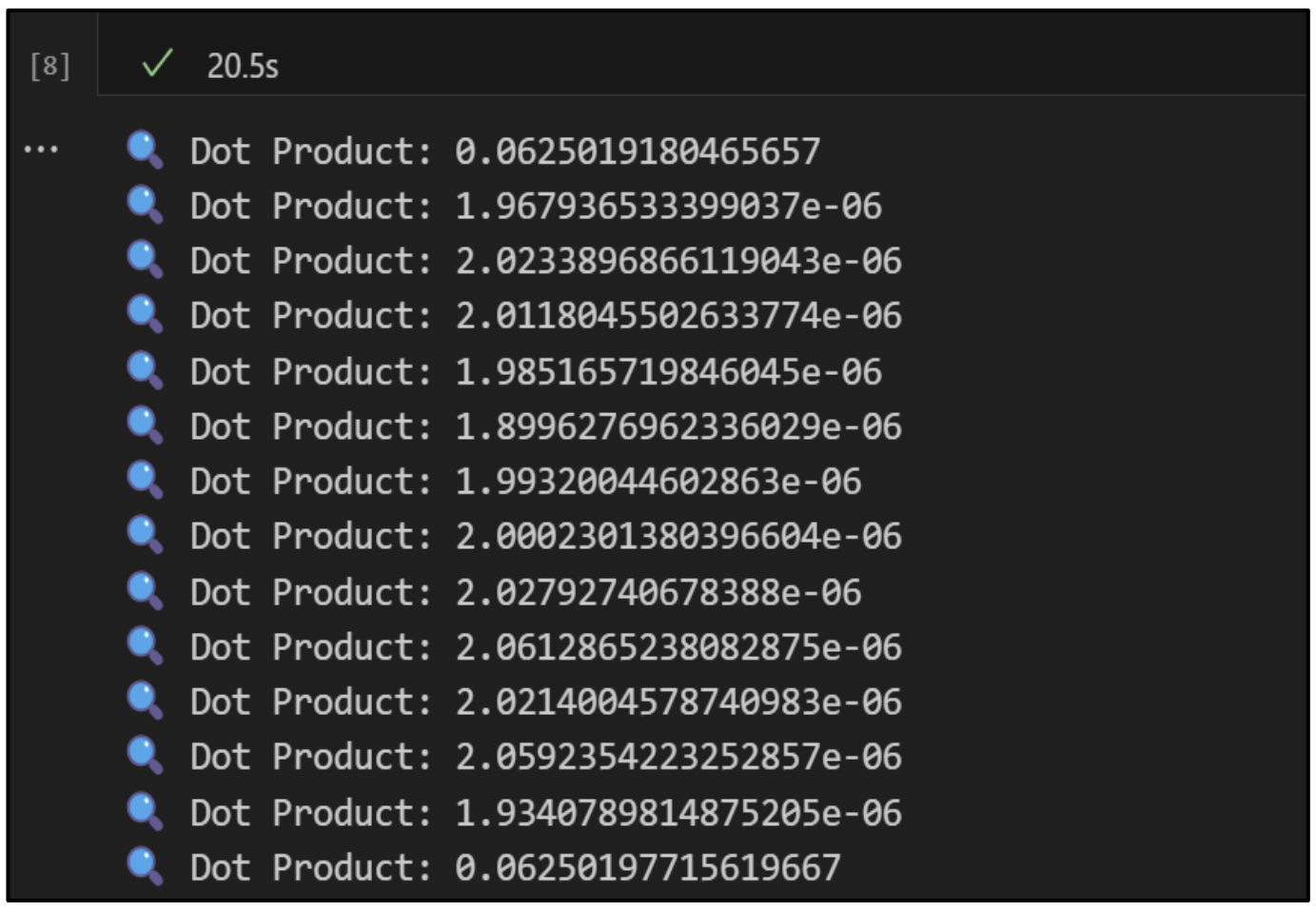

*Figure 4. Result of the query search "Identity Theft" when encrypted.*

Data Output Messages Notifications

Showing rows:

| | id [PK] integer | encryption_key_id integer | crime_type text | location text | description text | date text | time text | criminal_name text |
|---|---|---|---|---|---|---|---|---|
| 1 | 1 | 1 | Identity Theft | Portland | Bank account compromised through phishing email | 2024-02-23 | 14:35 | Unknown |
| 2 | 14 | 1 | Identity Theft | Detroit | Mail stolen from multiple residences | 2024-02-15 | 06:50 | Unknown |
| 3 | 35 | 1 | Identity Theft | Jacksonville | Victim's information used to file tax return | 2024-02-22 | Unknown | Unknown |
| 4 | 57 | 1 | Identity Theft | Orlando | Hotel rooms booked with stolen information | 2024-01-14 | 18:30 | Unknown |
| 5 | 71 | 1 | Identity Theft | Scottsdale | Medical services billed under victim's name | 2024-02-19 | 11:25 | Unknown |
| 6 | 83 | 1 | Identity Theft | Santa Ana | Multiple accounts opened using stolen identity | 2024-02-28 | Unknown | Unknown |

Total rows: 6 Query complete 00:00:00.058

*Figure 5. Result of the query search "Identity Theft" when stored as a plaintext.*

- Encryption / Decryption time

Encryption and Insertion Time:

Encrypting crime report entries utilized hybrid encryption: CKKS for searchable fields (such as crime_type) while other sensitive fields (i.e., location, description, criminal_name) were encrypted with BFV. The time taken for the operation was approximately 11.6 seconds which is depicted in the Figure 7. for a dataset of some 100 records undergoing encryption and database insertion over three federated nodes, wherein all vectorization and encryption algorithms were securely inserted into a database thereafter.

Decryption Time:

The decryption of the encrypted records during retrieval took about 2.1 seconds as shown in Figure 6, indicating that even though there was some overhead due to encryption, the decryption remained efficient and viable for interactive querying.

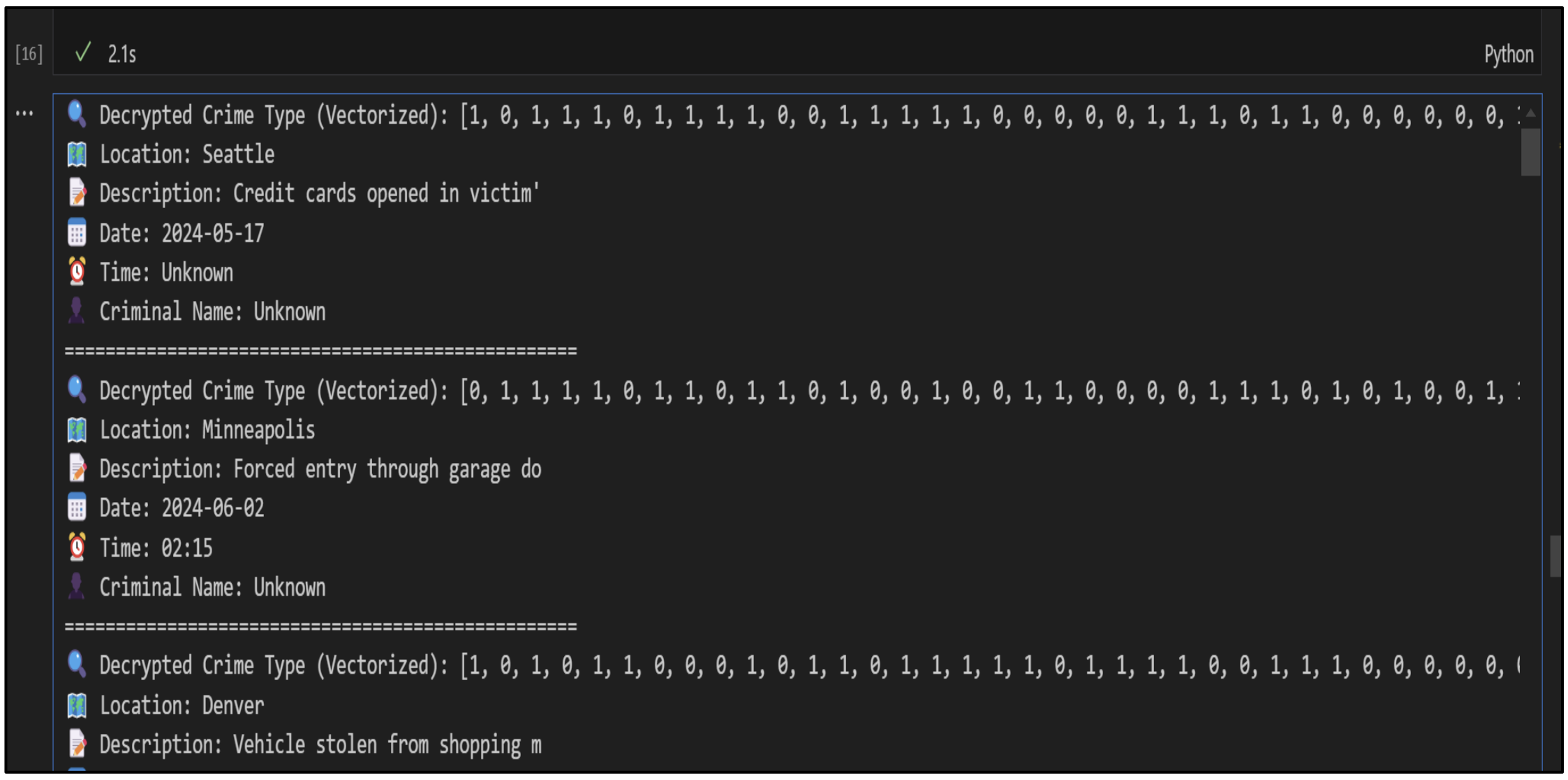

*Figure 6. The time taken to decrypt and display the data.*

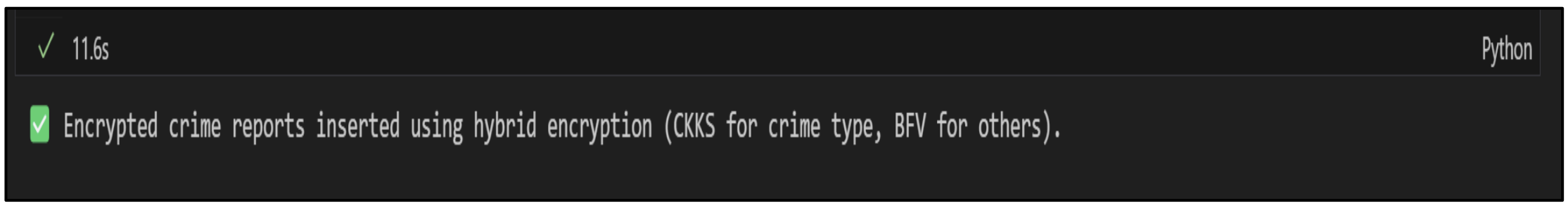

*Figure 7. Time taken to encrypt and store the data.*

- Search accuracy (true/false positives)

  Since only one column is taking part in the dot product operation (crime_type), using vectorized representations, this system gives a very high current accuracy of 100% in retrieving relevant results. Such impressive accuracy is mainly due to the very simple search process and the relatively small sizes of the dataset involved in it. However, as the dataset increases and gets diversely populated, the chances of accuracy falling become greater, especially when records contain more or less similar or overlapping keywords. Also, the fixed threshold limit that the system employs to determine similarity does not scale well into larger datasets and can lead to false positives or missed matches. However, further tests beyond what is presented here, on larger datasets are further to validate the robustness and scalability of the achieved search accuracy, even though current results are promising.

- Scalability (number of nodes, dataset size)

  The current system works on three federated databases, each with about 100 encrypted crime reports. While this setup proves the basic functionality very well, scaling would still be an important factor to consider when deploying in real-world scenarios. With more nodes, the search process becomes increasingly denigrated by the time taken to query and compute homomorphic operations over multiple databases in a sequential manner. Even going with larger datasets in the single central node, the processes of encryption, storage, and retrieval for CKKS-based dot product computations might also have frequently occurring peaks of latency. However, the design can be described as horizontally scalable--adding another database or node can take place with minimum changes to configuration. Future optimizations like parallel querying, load balancing, and adaptive thresholding can be potential candidates to improve performance and keep efficiency at scale.

## 6.5 Test Scenarios

- Search for criminal record type across agencies

To assess the search functionality of the system among federated agencies, a query was carried out using the crime type "Identity Theft." The system was able to successfully return relevant results with the matching of the encrypted vectorized crime type using homomorphic dot product comparison. The search was performed across the three federated databases and reached completion in approximately 16.7 seconds. This response time is way over that of a plaintext search and serves to show the ability of the system to maintain privacy and accurately fetch results in an encrypted domain. A screenshot of the retrieved data is provided in the Figure 8, which supports the validity of the system.

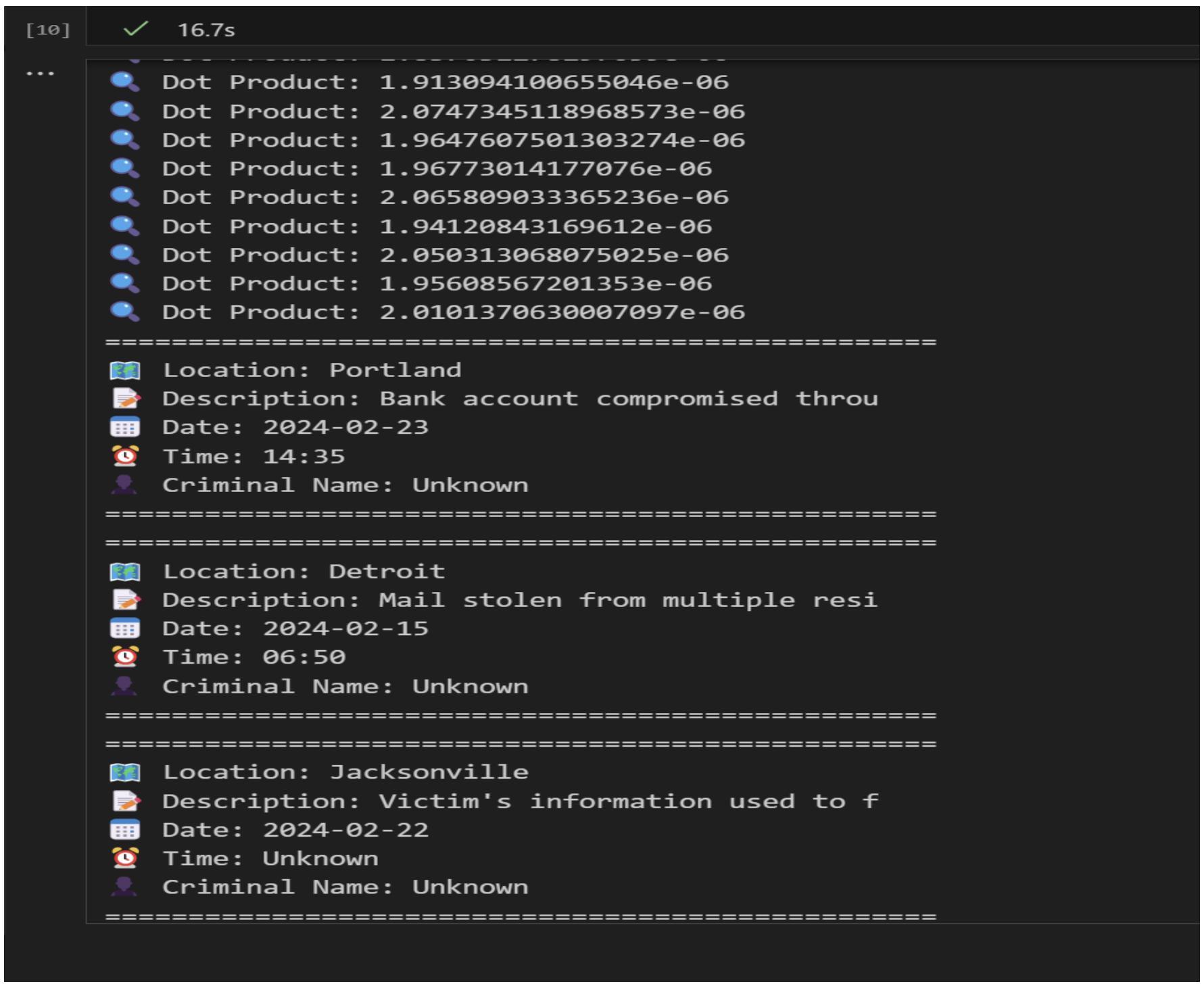

*Figure 8. Output of the query "Identity Theft"*

- Filter by attributes (location, date, criminal name)

This system can also perform post-query filtering through additional parameters such as location, date, and criminal name. Having executed the main search query, say find reports on Identity Theft, the System retrieves all relevant matches from its databases. The users can further narrow down the results by entering filters, such as the specific location ("Portland"), dates, or known criminal names like shown in the Figure 9. In essence, this approach yields cleaner and more meaningful results but still operates under the privacy-preserving aspect of the system. Indeed, even after a homomorphic search, further granularity may still be obtained without re-querying the encrypted databases.

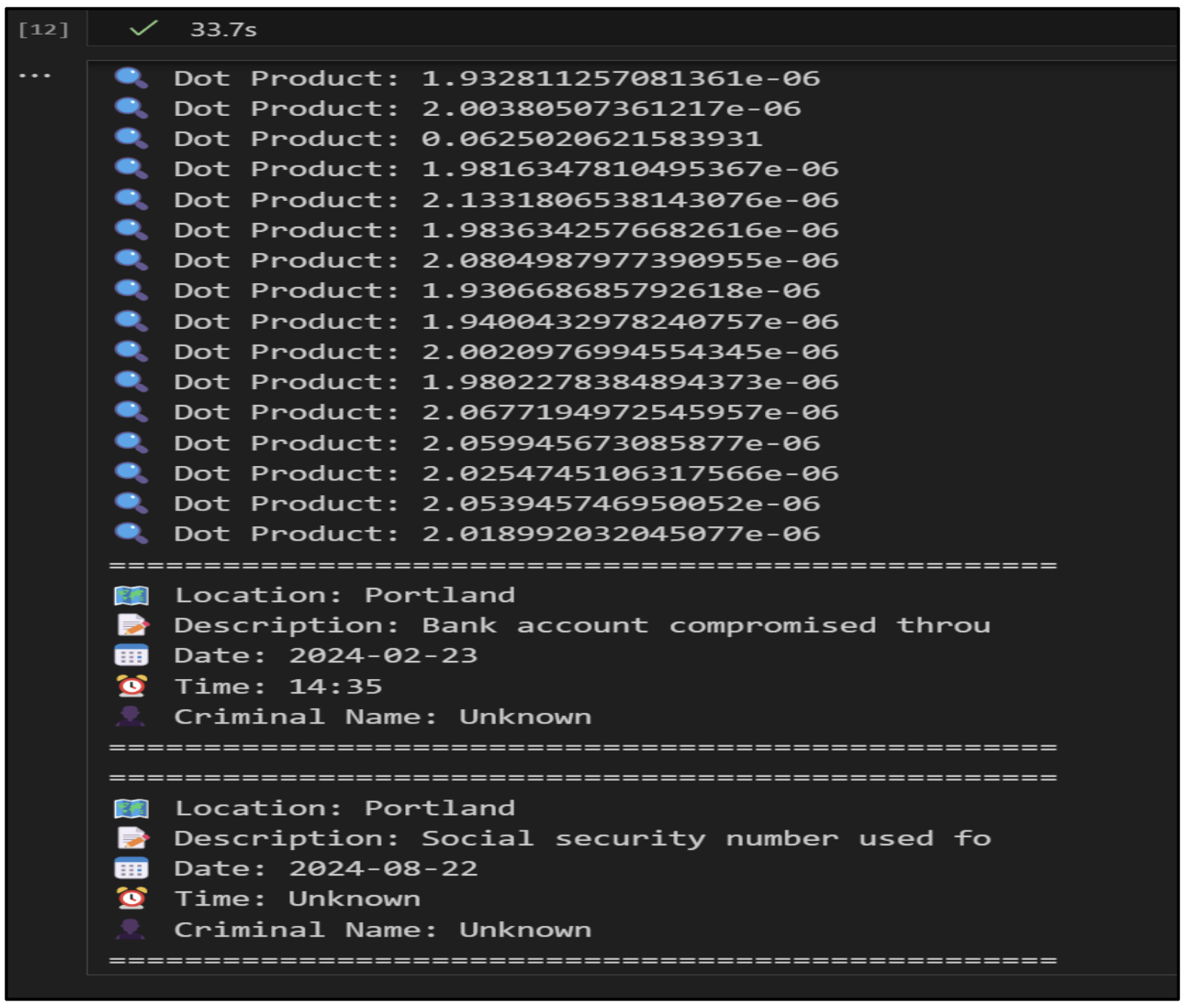

*Figure 9. Output of the query "Identity Theft" and applied filter of location "Portland".*

## 6.6 Observations

The trade-off between the speed of the operation and the extent of privacy has been well discussed in this implementation. Plaintext search has been extremely unadvised, taking about 0.058 seconds per database. Whereas, the homomorphic encryption-based dot product search is taking approximately 20.5 seconds so far for three databases. That inconvenience in the latency is what is considered data confidentiality and secure computation. It may not be favorable in performance, however, being able to search without decrypting data gives a stronger shelf on privacy, which will come in handy in sensitive applications such as inter-agency crime records sharing.

The federated architecture proved as a very efficient way to allocate encrypted data into more than one database. This not only decentralizes and makes the system strong against failure, but also enables various agencies to be operational independently, while the system has the possibility of secure collaborative querying. All databases have the same schema which ensures uniformity and hence facilitates encryption, storage, and searching across the system.

Performance drawbacks are generally always with encryption as well as decrypting, both of which are relatively efficient—around 11.6 seconds to perform encryption, while decryption takes around 2.1 seconds for the whole dataset. But the real delay is seen during the secure matching phase. It has a payload overhead introduced by CKKS vector operations prominently contributing to this delay. Subsequent iterations can optimize this by using caching for frequently used vectors, enhancing the hardware acceleration, or light cryptographic mechanisms for a particular query type.

Though t currently maintains 100% search accuracy on smaller datasets, it may not hold as the dataset grows; if the vector overlap becomes less distinct, the precision will be affected. Implementing adaptively thresholding or probabilistic vector encoding can keep the accuracy with larger numbers of datasets.

Finally, the post-query location, date, or by criminal name attribute-based filtering is introducing yet another very important facet of flexibility in this approach, allowing the users to further limit the search under security. Overall, while there is some latency with this approach, it gives great privacy, scalable architecture, and valuable effects on real-world secure search applications.

## Conclusions

To address the need for secure handling of data by law enforcement, managing privacy with inter-agency cooperation becomes an important challenge for agencies. Traditional systems either expose very sensitive information or are not amenable to meaningful cooperation. Our solution here is to enable encrypted, federated search across multiple databases while retaining data confidentiality and allowing useful querying.

Flying this hybrid encryption model is a combination of homomorphic encryption schemes: CKKS for searchable fields and BFV for sensitive string data. A dot product matching on encrypted vectors prevents any leakage of sensitive information during search and provides a federated architectural model allowing each agency to keep control of their datasets while joining in the alliance search queries.

The system solves an immediate problem: how to implement secure data sharing across multiple agencies. However, it offers much broader long-term promise in cross-border policing and secure intelligence exchange. The ability to collaborate on criminal investigations without compromising individual or national privacy norms becomes significant in an increasingly interconnected, digital, and globalized world. We have built a good foundation for scalable and secure, private crime data collaborations, promising new paths for the future of law enforcement and secure data science.

# Code and data Availability:

Google Colab Link